\documentclass[a4,14pt]{article}
\usepackage[margin=1in]{geometry}
\usepackage{subcaption}
\usepackage{cite}
\usepackage{amsmath}
\usepackage{hyperref}
\usepackage{amssymb}
\usepackage{slashed}
\usepackage{braket}
\usepackage{upgreek}
\usepackage{graphicx}
\usepackage{float}
\usepackage{slashed}
\begin{document}
\title{Screening models and neutrino oscillations}
\author{H. Yazdani Ahmadabadi\footnote{hossein.yazdani@ut.ac.ir} and H. Mohseni Sadjadi\footnote{mohsenisad@ut.ac.ir}
\\ {\small Department of Physics, University of Tehran,}
\\ {\small P. O. B. 14395-547, Tehran 14399-55961, Iran}}
\maketitle
\begin{abstract}
In screening models with scalar-matter conformal coupling, we study the flavor transition of neutrinos.
We employ an analytical method for studying the oscillation phase in a spherically symmetric spacetime filled by a scalar field.
Since the ambient matter density determines the scalar field's behavior, an indirect environmental effect contributes to the flavor conversion inside matter.
We evaluate the survival probabilities and show that the existence of the scalar field affects the oscillations of neutrinos.
We discuss the results in the framework of screening mechanisms and the end, confront our results with observational data.
\end{abstract}

\section{Introduction}\label{sec1}
One of the interesting problems in particle physics is the neutrino's flavor oscillation which implies that neutrinos have non-zero masses.
Flavor oscillation was first proposed in \cite{Pontecorvo:1957cp} to explain the solar neutrino problem \cite{Sun} and later experimentally verified by detectors in Super-Kamiokande (SK) \cite{Super-Kamiokande:2002weg} and the Sudbury Neutrino Observatory (SNO) \cite{SNO:2002tuh}.
Also, several experiments using solar \cite{SNO+SK.Solar1,SNO+SK.Solar2,SNO+SK.Solar3,SNO+SK.Solar4}, atmospheric \cite{IceCube+SK.Atm1,IceCube+SK.Atm2,IceCube+SK.Atm3}, reactor \cite{DayaBay+RENO.RAC1,DayaBay+RENO.RAC2}, and accelerator \cite{MINOS.ACC} neutrinos have been done, and data analyzes have been performed to calculate the neutrino mass and mixing parameters.
To explain the deficit in the number of solar and atmospheric neutrinos \cite{mu1,mu2,mu3}, it is accepted that neutrinos flavors change among three flavors considered mixtures of the mass states.
When neutrinos pass through matter, e.g., in the sun, a subtle and significant phenomenon occurs; they experience forward scattering from solar matter electrons, and the MSW (Mikheyev-Smirnov-Wolfenstein) effect \cite{MSWWolf,MSWMS} happens, which is an adiabatic flavor conversion in a medium with slowly mass varying density.

Apart from the neutrino oscillations in the flat spacetime, neutrino flavor conversion in curved spacetime\cite{Stodolsky} has attracted much attention in the extended theories of gravity \cite{Curved1,Curved.Fornengo,Curved2.Visinelli,Curved3,Curved4,Curved5,Curved6,Curved7.Sadjadi,sadeg,kalif}.
The interaction of neutrinos with a scalar field and its cosmological impacts have been studied vastly in the literature \cite{impac1,impac2,impac3,impac4,impac5}.
The interaction of a scalar field and neutrinos through a conformal coupling and its influence on the neutrino density and their oscillation phases were discussed in \cite{Curved7.Sadjadi,new1,new2,new3}.
This kind of conformal couplings has also been used to describe the onset of the late cosmic acceleration when the neutrinos became non-relativistic\cite{anari1,anari2}, and describe an early cosmic acceleration in the matter-radiation equality epoch to solve the Hubble tension problem\cite{trod}.

Also, the modified MSW effect caused by the curvature of spacetime has been discussed in \cite{Zhang}.
In some kinds of scalar-tensor dark energy theories, the quintessence, which interacts with matter through a conformal coupling, can be a source of curvature of spacetime \cite{mat1,mat2,mat3}.
This coupling results in the screening effect, e.g., in the Chameleon \cite{Khoury:2003aq,Khoury:2003rn,Waterhouse:2006wv,Tsujikawa:2009yf,Burrage:2017qrf} or the Symmetron\cite{Hinterbichler:2010es,Hinterbichler:2011ca,sad3,Honardoost:2017hpj} models.

The present paper explores the indirect influence of matter on the neutrino flavor change process through a scalar field that is coupled to both the matter and neutrinos.
In section \ref{sec2}, we analyze the covariant formulation of the phase shift in the vacuum.
Corrections to the neutrino quantum mechanical phase difference and oscillation probability related to the curved spacetime and in the presence of a scalar field are explicitly calculated by the WKB approach in this section.
The explicit forms of the oscillation phase in the presence of the Chameleon and Symmetron scalar fields are obtained in section \ref{sec3}.
Modified weak interaction with electrons inside the solar matter, which affects the flavor conversion process is presented in section \ref{sec4}.
We finally summarize our results in section \ref{sec5} through some numerical examples.
The paper's results will be compared with the SNO + SK \cite{Zyla:2020zbs} and the Borexino \cite{Agostini:2018uly} data for the MSW-LMA (Large Mixing Angle) survival probability.
The detailed calculations for screening mechanisms are also provided in Appendix \ref{app1}: Appendix \ref{subapp1}, which contains all calculations for the Chameleon scalar field, and the treatment of the Symmetron field is presented in Appendix \ref{subapp2}.

Throughout this paper we use units $\hbar=c=1$ and metric signature $(-,+,+,+)$.

\section{The Dirac equation and the phase difference}\label{sec2}
In our model, the metric in the matter sector is conformally coupled to a scalar field $\phi$.
We consider the action
\begin{eqnarray}\label{eqn1}
S = \int d^4 x \sqrt{-\text{g}}\left[\frac{M_p^{2}}{2}\mathcal{R} - \frac{1}{2}\text{g}^{\mu\nu}\partial_\mu\phi \partial_\nu\phi - V(\phi) \right] + \int d^4x \mathcal{L}_\text{m} \left(\Psi_{i},\tilde{\text{g}}^{(i)}_{\mu\nu}\right),
\end{eqnarray}
where $\mathcal{R}$ is the scalar curvature, $M_p \equiv (8\pi G)^{-1/2}$ is the reduced Planck mass, $\Psi_{i}$ is the $i$th matter field, and $\mathcal{L}_\text{m}$ is the matter Lagrangian density.
$\tilde{\text{g}}^{(i)}_{\mu\nu}$ is given by  \cite{Faraoni,Shapiro:2016pfm}:
\begin{eqnarray}\label{eqn2}
\tilde{\text{g}}^{(i)}_{\mu\nu} = A_i^2\left(\phi\right) \text{g}_{\mu\nu}.
\end{eqnarray}
$A_i\left(\phi\right)$ is the conformal factor.
The scalar field potential is given by $V(\phi)$.
The screening models do not satisfy the weak equivalence principle (WEP) unless we take a universal coupling function for all matter species $\Psi_i$'s \cite{Burrage:2017qrf}.
For the Schwarzschild metric, we have:
\begin{eqnarray}\label{eqn3}
\begin{split}
ds^2_{(i)} = - A_i^2\left(\phi\right)\left(1- \frac{2 G m(r)}{r}\right)dt^2 + A_i^2\left(\phi\right)\left(1- \frac{2Gm(r)}{r}\right)^{-1} dr^2 + r^2 A_i^2\left(\phi\right) \left(d\theta^2 + \sin^2\theta~d\upvarphi^2\right),
\end{split}
\end{eqnarray}
where
\begin{eqnarray}\label{eqn4}
m(r) = \int_{0}^{r} 4\pi r^{\prime 2} \rho(r^\prime) dr^\prime
\end{eqnarray}
is the mass and $\rho(r)$ is the mass density of the object.
In this section, we use the WKB approximation \cite{Curved.Fornengo,Curved2.Visinelli}, in which the wavefunction of the massive neutrino is decomposed into a semi-classical phase and an amplitude.
We solve the Dirac equation for a spacetime described by the line element (\ref{eqn3}), and the phase shift of vacuum oscillations (oscillations outside the spherical object) can then be obtained straightforwardly.
By restoring $\hbar$, the Dirac equation is then written as
\begin{eqnarray}\label{eqn5}
\left(i\hbar\gamma^\mu D_\mu - m_\nu \right) \nu(x)=0,
\end{eqnarray}
where $D_\mu = \partial_\mu + \Gamma_\mu$.
Moreover, the spin connection is given by
\begin{eqnarray}\label{eqn6}
\Gamma_\mu = \frac{1}{8}[\gamma^{\hat{a}} , \gamma^{\hat{b}}] e_{\hat{a}}^\nu e_{{\hat{b}} \nu;\mu}
\end{eqnarray}
where the tetrads are denoted by $e^{\hat{a}}_\mu$.
We have \cite{Curved1}
\begin{eqnarray}\label{eqn7}
\gamma^{\hat{a}} e^\mu_{\hat{a}} \Gamma_\mu = \gamma^{\hat{a}} e^\mu_{\hat{a}} \left\{i A_{G\mu} \left[-\left(-\tilde{\text{g}}^{(i)}\right)^{-\frac{1}{2}} \frac{\gamma_5}{2}\right]\right\},
\end{eqnarray}
where
\begin{eqnarray}\label{eqn8}
A^\mu_G \equiv \frac{1}{4} \left(-\tilde{\text{g}}^{(i)}\right)^{-\frac{1}{2}} e_{\hat{a}}^\mu \epsilon^{\hat{a}\hat{b}\hat{c}\hat{d}} \left(e_{\hat{b}\nu,\sigma} - e_{\hat{b}\sigma,\nu}\right) e^\nu_{\hat{c}} e^\sigma_{\hat{d}}.
\end{eqnarray}
Here, $\tilde{\text{g}}^{(i)}$ is the determinant of the metric Eq.(\ref{eqn2}), and $\epsilon^{\hat{a}\hat{b}\hat{c}\hat{d}}$ is the totally antisymmetric Levi-Civita symbol.
As we can expect from diagonal metric in Eq.(\ref{eqn3}), $A^\mu_G = 0$ \cite{Curved1,Curved6,Piriz:1996mu}.
We can, of course, conclude this from direct calculations by using the diagonal tetrad field
\begin{eqnarray}\label{eqn9}
e^{\hat{a}}_\mu = \text{diag} \left\{A(\phi) \left(1-\frac{G m(r)}{r}\right), A(\phi) \left(1+\frac{G m(r)}{r}\right), A(\phi) r, A(\phi) r \sin\theta\right\}.
\end{eqnarray}
In fact, a simple analysis of Eq.(\ref{eqn8}) shows that a non-vanishing $A_G^\mu$ requires non-zero off-diagonal elements of the tetrads \cite{Curved6}.
Furthermore, tetrads satisfy the relation $\text{g}_{\mu\nu} = e^{\hat{a}}_{\mu} e^{\hat{b}}_{\nu} \eta_{\hat{a}\hat{b}}$, where the Greek letters $\mu,\nu = \{0,1,2,3\}$ denote indices related to the spacetime manifold, and the Latin alphabet with a hat $\hat{a},\hat{b}$ denote the indices related to the locally inertial coordinates.
We are now going to solve the Dirac equation (\ref{eqn5}) to compute the vacuum phase of oscillations.

Following Refs.\cite{Curved.Fornengo,Curved2.Visinelli}, we expand the wave function $\nu(x)$ in powers of $\hbar$, i.e., we use
\begin{eqnarray}\label{eqn10}
\nu(x) = e^{-i\varphi(x)/\hbar} e^{-\Gamma_\mu x^\mu} \sum_{k=0}^{\infty} \left(\frac{\hbar}{i}\right)^k \nu_k(x),
\end{eqnarray}
where $\varphi(x)$ is the semi-classical phase, and $\nu_k$'s are the amplitudes of the spinor for different $k$'s.
Substituting (\ref{eqn10}) in the Dirac equation (\ref{eqn5}), the two below relations are obtained:
\begin{eqnarray}\label{eqn11}
\begin{cases}
(\slashed{\partial} \varphi - m)\nu_0 = 0~~~~~~~~~~~~~~~~~~~~~(k=0)\\
(\slashed{\partial} \varphi - m)\nu_k = \slashed{\partial} \nu_{k-1}~~~~~~~~~~~~~~~(k \neq 0)
\end{cases}
.
\end{eqnarray}
Multiplying the first relation from left by $(\slashed{\partial}\varphi + m)$ leads to the Hamilton-Jacobi equation
\begin{eqnarray}\label{eqn12}
\partial_\mu \varphi(x) \partial^\mu \varphi(x) = m^2.
\end{eqnarray}
We obtain the following phase of oscillations by comparing Eq.(\ref{eqn12}) to the mass-shell condition $- p_\mu p^\mu= m^2$ for a particle with mass $m$ and momentum $p$:
\begin{eqnarray}\label{eqn13}
\varphi(x) = - \int^{x}_{x_0} p_\mu dx^{\mu}.
\end{eqnarray}
This is the familiar covariant form of the phase function of the spinor $\nu(x)$ produced at the spacetime point $x_0$ and detected at the point $x$. Also, $E \equiv - p_0 = -\tilde{\text{g}}_{0 \mu} p^\mu$.

We now obtain a specific expression for the phase $\varphi(x)$ in our model.
As it can be seen, the line element (\ref{eqn3}) is symmetric under changes of the coordinates $t$ and $\upvarphi$.
Thus, because of the isometry, this metric admits two Killing vectors, $\partial_t$, and $\partial_\upvarphi$.
The energy $E$ and azimuthal angular momentum $L$ are therefore both conserved in this spacetime.
The phase (\ref{eqn13}) is then written as
\begin{eqnarray}\label{eqn14}
\varphi = E(t-t_0) - L(\upvarphi-\upvarphi_0) - \mathcal{P}(r) - \mathcal{L}(\theta),
\end{eqnarray}
where $\mathcal{P}(r)$ and $\mathcal{L}(\theta)$ are respectively the radial and polar-angle contributions in the above phase.
Using Eq.(\ref{eqn12}) and inserting the elements of the inverse metric gives
\begin{eqnarray}\label{eqn15}
\left(1 - \frac{r_g}{r}\right)^{-1} E^2 - \left(1 - \frac{r_g}{r}\right) \left(\frac{d\mathcal{P}(r)}{dr}\right)^2 -\frac{1}{r^2} \left(\frac{d\mathcal{L}(\theta)}{d\theta}\right)^2 - \frac{L^2}{r^2 \sin^2\theta} = - m^2 A^2(\phi),
\end{eqnarray}
where, for the Sun,  $r_g \equiv 2G M_{\odot}$ is the Schwarzschild radius and $M_{\odot}$ is the mass.
By using the separation of variables, we have two equations
\begin{eqnarray}\label{eqn16}
\left(\frac{d\mathcal{L}(\theta)}{d\theta}\right)^2 + \frac{L^2}{\sin^2\theta}=K,
\end{eqnarray}
and
\begin{eqnarray}\label{eqn17}
\left(\frac{d\mathcal{P}(r)}{dr}\right)^2= \left(1 - \frac{r_g}{r}\right)^{-2} E^2 + m^2 A^2(\phi) \left(1 - \frac{r_g}{r}\right)^{-1} -\frac{K}{r^2} \left(1 - \frac{r_g}{r}\right)^{-1},
\end{eqnarray}
where we have assumed that $r_g/r$ is small enough to keep only the first-order terms in gravitational potential expansion.
In the case of  propagation of neutrinos in a special plane with a fixed angle $\theta_0$,  $K = L^2/ \sin^2 \theta_0$, which means that $\mathcal{L}(\theta_0)=cte$.
For the sake of simplicity, this constant is considered to be zero.

Beside, the relation for $\mathcal{P}(r)$ is obtained by
\begin{eqnarray}\label{eqn18}
\mathcal{P}(r)= \int^r_{r_0} dr^\prime \sqrt{\left(1 - \frac{r_g}{r^\prime}\right)^{-2} E^2 + m^2 A^2 \left(1 - \frac{r_g}{r^\prime}\right)^{-1}},
\end{eqnarray}
and hence the oscillation phase (\ref{eqn14}) for various scalar-matter couplings becomes
\begin{eqnarray}\label{eqn19}
\varphi_i (r,t) = E(t-t_0)-\int^r_{r_0} dr^\prime \sqrt{\left(1 - \frac{r_g}{r^\prime}\right)^{-2} E^2 + m_i^2 A_i^2 \left(1 - \frac{r_g}{r^\prime}\right)^{-1}}.
\end{eqnarray}
As a particular case, we have only considered the radial propagation, i.e., $K=L=0$.
In the final stage, we Taylor-expand the oscillation phase in terms of $m^2_i$ as follows \cite{Curved2.Visinelli}
\begin{eqnarray}\label{eqn20}
\varphi(m_i,r_B - r_A)=\sum_{k=0}^{+\infty}\frac{(m_i^2)^k}{k!} \varphi_i^{(k)}(r_B - r_A),
\end{eqnarray}
where $r_B - r_A$ is the coordinate distance between the source and detector.
By taking derivatives with respect to $m^2_i$, two first terms of this expansion can be obtained.
We have
\begin{eqnarray}\label{eqn21}
\varphi^{(0)}_i(r,t)=E(t-t_0)-E \int^r_{r_0} dr^\prime \left(1 + \frac{r_g}{r^\prime}\right),
\end{eqnarray}
\begin{eqnarray}\label{eqn22}
\varphi^{(1)}_i(r)=\left(\frac{\partial \varphi_i(r,t)}{\partial m_i^2}\right)_{m_i^2=0} = - \frac{1}{2E}\int^r_{r_0} dr^\prime A_i^2(\phi),
\end{eqnarray}
and
\begin{eqnarray}\label{eqn23}
\varphi_i^{(2)}(r)=\left(\frac{\partial^{(2)} \varphi_i(r,t)}{\partial {m_i^2}^{(2)}}\right)_{m_i^2=0} = \frac{1}{4E^3} \int^r_{r_0} dr^\prime A_i^4(\phi) \left(1 - \frac{r_g}{r^\prime}\right),
\end{eqnarray}
where the relation for the zeroth-order term is computed by putting $m_i^2=0$ in Eq.(\ref{eqn19}).
We know that the scalar field $\phi$ in the above relation is an explicit function of the radial coordinate (see the Appendix \ref{app1}).
The phase difference between various mass eigenstates of neutrinos propagating from point $r_A$ to $r_B$ is then given by
\begin{eqnarray}\label{eqn24}
\begin{split}
&\Phi_{ij} = - \frac{1}{2E} \left[m^2_i \int_{r_A}^{r_B} dr A_i^2(\phi) - m^2_j \int_{r_A}^{r_B} dr A_j^2(\phi)\right] \\&~~~~~~ + \frac{1}{8E^3} \left[m_i^4 \int_{r_A}^{r_B} dr A_i^4(\phi)\left(1 -\frac{r_g}{r}\right) - m_j^4 \int_{r_A}^{r_B} dr A_j^4(\phi)\left(1 -\frac{r_g}{r}\right)\right].
\end{split}
\end{eqnarray}
The second term is due to the gravitational potential and scalar-matter coupling, whereas the first term originates only from the conformal coupling function.
This result is the same as in Ref.\cite{Curved7.Sadjadi}, provided that we neglect the term proportional to $\mathcal{O}(m^4/E^3)$.
It is also worth noting that, to respect the WEP and simplify the numerical estimations, we will consider a universal coupling parameter $\beta$ for all matter species in the following sections.

\section{Neutrino flavor transition}\label{sec3}
Neutrino flavor oscillation is a process when a neutrino is created with a particular flavor $\alpha$ in the source and then travels a relatively long distance and finally contributes to weak interactions in the detectors to show a neutrino with a different flavor $\beta$.
This phenomenon has led to a change in our belief in the mass of neutrinos.
Here, we are interested in finding the expression for the neutrino oscillation phase in a spacetime filled with a scalar field.

In a standard treatment, the $\alpha$-neutrino state at the spacetime point $(t,r)$ is given by
\begin{eqnarray}\label{eqn25}
\ket{\upnu(r,t)}_\alpha = \sum_{i=1}^{3} U_{\alpha i} e^{-i\varphi_i(r,t)} \ket{\nu_i},
\end{eqnarray}
where $U_{\alpha i}$'s are the elements of a $3 \times 3$ unitary matrix called PMNS (Pontecorvo-Maki-Nakagawa-Sakata) matrix, which comprises the notion of mixing, and $\varphi_i(r,t) = - \int^x p_\mu dx^\mu$ is the phase that illustrates the spacetime evolution of the $i$th mass eigenstate.
The probability of neutrino flavor change to $\beta$ can then be written as follows:
\begin{eqnarray}\label{eqn26}
\begin{split}
&P_{\alpha \beta} = \delta_{\alpha\beta} -4 \sum_{i>j} \Re\left(U_{\alpha i} U_{\beta i}^* U_{\alpha j}^* U_{\beta j}\right) \sin^2\left(\frac{\Phi_{ij}}{2}\right) \\& ~~~~~+ 2\sum_{i>j} \Im \left(U_{\alpha i} U_{\beta i}^* U_{\alpha j}^* U_{\beta j}\right) \sin\left(\Phi_{ij}\right),
\end{split}
\end{eqnarray}
where $\Phi_{ij} \equiv \varphi_i - \varphi_j$ is the phase difference.
To calculate the probabilities directly, we need to look at the unitary mixing matrix $U$.
It is illuminating to write the matrix in the form
\begin{eqnarray}\label{eqn27}
\begin{split}
&U =
\begin{pmatrix}
c_{12} c_{13} & s_{12} c_{13} & s_{13}e^{-i\delta}\\
-s_{12} c_{23} - c_{12} s_{23} s_{13} e^{i\delta} & c_{12} c_{23} - s_{12} s_{23} s_{13} e^{i\delta} & s_{23} c_{13}\\
s_{12} s_{23} - c_{12} c_{23} s_{13} e^{i\delta} & -c_{12} s_{23} -s_{12} c_{23} s_{13} e^{i\delta} & c_{23} c_{13} \\
\end{pmatrix}
,
\end{split}
\end{eqnarray}
where $c_{ij}\equiv \cos\theta_{ij}$, $s_{ij}\equiv\sin\theta_{ij}$, and $\delta$ is the Dirac CP-violating phase.
As the simplest case, the transition probability for two flavors $\{e,\mu\}$ is given by
\begin{eqnarray}\label{eqn28}
P_{\alpha \beta} = \delta_{\alpha \beta} - \left(2\delta_{\alpha
\beta} -1 \right) \sin^2 (2\theta) \sin^2\left(\frac{\Phi_{12}}{2}\right).
\end{eqnarray}

The scalar field has been considered in the Appendix \ref{app1} to be expanded around its minimum value, i.e., we have $\phi_{\text{out}}(r) = \phi_0 + \delta \phi_{\text{out}}(r)$.
Substituting this in the phase of oscillations in the presence of the Chameleon scalar field leads to
\begin{eqnarray}\label{eqn29}
\begin{split}
&\varphi_i \simeq -E r_g \ln\left(\frac{r_B}{r_A}\right) - \frac{m_i^2}{2E} e^{\frac{2\beta \phi_0}{M_p}} \left[|r_B-r_A| + \frac{2\beta}{M_p} \int_{r_A}^{r_B} dr~\delta \phi_{\text{out}}(r)\right] \\& ~~~ + \frac{m_i^4}{8E^3} e^{\frac{4\beta \phi_0}{M_p}} \left[|r_B-r_A| - r_g \ln \left(\frac{r_B}{r_A}\right) + \frac{4\beta}{M_p} \int_{r_A}^{r_B} dr~\delta\phi_{\text{out}}(r) - \frac{4\beta r_g}{M_p} \int_{r_A}^{r_B} dr~\frac{\delta \phi_{\text{out}}(r)}{r}\right].
\end{split}
\end{eqnarray}

To have more physical cases, we represent the phase difference Eq.(\ref{eqn24}) in terms of two physical quantities: Local energy $E_l(r_B)$ (measured by an observer on the detector at $r_B$) and proper distance $L_p$ (measured by an observer of the neutrino rest frame).
Local energy is defined by
\begin{eqnarray}\label{eqn30}
E_l(r_B) = \frac{E}{\sqrt{-g_{tt}(r_B)}} = \frac{E}{A\left[\phi(r_B)\right] \left(1-\frac{r_g}{r_B}\right)^{\frac{1}{2}}},
\end{eqnarray}
where $E$ is the energy of neutrino measured by an observer at infinity, and $g_{tt}(r_B)$ is the timelike component of the metric.
Since the coordinate distance is not a physical quantity in the curved spacetime, this will be replaced by the proper distance.
The proper distance is defined by
\begin{eqnarray}\label{eqn31}
L_p = \int_{r_A}^{r_B} \sqrt{g_{rr}}~dr \simeq \int^{r_B}_{r_A} A(\phi) \left(1+\frac{r_g}{2r}\right) dr,
\end{eqnarray}
where $g_{rr}$ is the radial component of the metric.
Using the definitions (\ref{eqn30}) and (\ref{eqn31}), we write the coordinate distance
\begin{eqnarray}\label{eqn32}
|r_B-r_A| = e^{-\frac{\beta \phi_0}{M_p}} L_p - \frac{r_g}{2} \ln \left(\frac{r_B}{r_A}\right) - \frac{\beta}{M_p} \int_{r_A}^{r_B} dr~\delta\phi_{\text{out}}(r) - \frac{\beta r_g}{2M_p} \int_{r_A}^{r_B} dr \frac{\delta\phi_{\text{out}}(r)}{r},
\end{eqnarray}
and energy
\begin{eqnarray}\label{eqn33}
E = E_l(r_B) \left(1-\frac{r_g}{r_B}\right)^{\frac{1}{2}} e^{\frac{\beta \left[\phi_0 + \delta\phi_{\text{out}}(r_B)\right]}{M_p}}.
\end{eqnarray}
By assuming $r_g/r_B \ll 1$ and substituting these relations in the phase of the oscillation (\ref{eqn29}), we obtain
\begin{eqnarray}\label{eqn34}
\begin{split}
& \varphi_i = -E_l(r_B) r_g e^{\frac{\beta \left[\phi_0 + \delta\phi_{\text{out}}(r_B)\right]}{M_p}} \ln\left(\frac{r_B}{r_A}\right) \\& ~~~ - \frac{m_i^2}{2E_l(r_B)} e^{\frac{\beta \left[\phi_0 - \delta\phi_{\text{out}}(r_B)\right]}{M_p}} \bigg[e^{-\frac{\beta \phi_0}{M_p}} L_p - \frac{r_g}{2} \ln\left(\frac{r_B}{r_A}\right) \\& ~~~+ \frac{\beta}{M_p} \int_{r_A}^{r_B}dr\delta\phi_{\text{out}}(r) - \frac{\beta r_g}{2M_p} \int_{r_A}^{r_B} dr \frac{\delta\phi_{\text{out}}(r)}{r} \bigg] \\& ~~~+ \frac{m_i^4}{8E^3_l(r_B)} e^{\frac{\beta \left[\phi_0 - 3\delta\phi_{\text{out}}(r_B)\right]}{M_p}} \bigg[e^{-\frac{\beta \phi_0}{M_p}} L_p - \frac{3r_g}{2} \ln\left(\frac{r_B}{r_A}\right) \\& ~~~+ \frac{3\beta}{M_p} \int_{r_A}^{r_B}dr\delta\phi_{\text{out}}(r) - \frac{9\beta r_g}{2M_p} \int_{r_A}^{r_B} dr\frac{\delta\phi_{\text{out}}(r)}{r} \bigg].
\end{split}
\end{eqnarray}
By ignoring the scalar-matter coupling, one obtains the oscillation phase in the Schwarzschild spacetime
\begin{eqnarray}\label{eqn35}
\begin{split}
&\varphi_i = -E_l(r_B) r_g \ln\left(\frac{r_B}{r_A}\right) - \frac{m_i^2}{2E_l(r_B)} \left[L_p - \frac{r_g}{2} \ln\left(\frac{r_B}{r_A}\right) \right] + \frac{m_i^4}{8E^3_l(r_B)} \left[ L_p - \frac{3r_g}{2} \ln\left(\frac{r_B}{r_A}\right) \right].
\end{split}
\end{eqnarray}

Repeating all steps above for the Symmetron scalar field, the phase of oscillations can be derived:
\begin{eqnarray}\label{eqn36}
\begin{split}
& \varphi_i = -E_l(r_B) r_g \left[1+ \frac{\phi_0^2}{2M^2} + \frac{\phi_0 \delta\phi_{\text{out}}(r_B)}{M^2}\right]\ln \left(\frac{r_B}{r_A}\right) \\& ~~~ - \frac{m_i^2}{2E_l(r_B)} \bigg[\left(1 - \frac{\phi_0 \delta\phi_{\text{out}}(r_B)}{M^2}\right) L_p - \left(1 + \frac{\phi_0^2}{2M^2} - \frac{\phi_0 \delta \phi_{\text{out}}(r_B)}{M^2}\right) \frac{r_g}{2} \ln\left(\frac{r_B}{r_A}\right) \\& ~~~+ \frac{\phi_0}{M^2} \int_{r_A}^{r_B} dr \delta\phi_{\text{out}}(r) - \frac{r_g \phi_0}{2M^2} \int_{r_A}^{r_B} dr \frac{\delta \phi_{\text{out}}(r)}{r} \bigg] \\& ~~~ +  \frac{m_i^4}{8 E^3_l(r_B)} \bigg[\left(1 - \frac{3\phi_0 \delta\phi_{\text{out}}(r_B)}{M^2}\right) L_p - \left(1 + \frac{\phi_0^2}{2M^2} - \frac{3\phi_0 \delta\phi_{\text{out}}(r_B)}{M^2}\right) \frac{3r_g}{2} \ln\left(\frac{r_B}{r_A}\right) \\& ~~~+ \frac{3\phi_0}{M^2} \int_{r_A}^{r_B} dr \delta\phi_{\text{out}}(r) - \frac{9r_g \phi_0}{2M^2} \int_{r_A}^{r_B} dr \frac{\delta \phi_{\text{out}}(r)}{r} \bigg].
\end{split}
\end{eqnarray}
In the above relation, we considered the neutrino propagation outside the object.
For inside, the matter MSW effect must also be considered, which is done in the next section.

\section{Neutrino propagation inside matter}\label{sec4}
We now discuss the matter effects on neutrinos flavor change inside a spherical star.
Neutrinos encounter numerous electrons and interact with them through the weak charged current, which affects the flavor change process as predicted decades ago by Mikheyev, Smirnov, and Wolfenstein, dubbed the MSW effect \cite{MSWWolf,MSWMS}.
The flavor conversion process is also modified due to the terms coming from the gravitational corrections of the neutrino self-energy, discussed in detail in \cite{Zhang}.

For simplicity, we address a simplified two-flavor (e.g., $\nu_e$ and $\nu_\mu$) model inside matter in the curved spacetime.
If the number density of electrons inside the star is a constant or even changes adiabatically and the scalar field can be considered approximately uniform, we can write the 2$\nu$ Schr$\ddot{\text{o}}$dinger-like evolution equation
\begin{eqnarray}\label{eqn37}
i \frac{d}{dl}
\begin{pmatrix}
\nu_e \\
\nu_\mu
\end{pmatrix}
= \frac{1}{4E_l(r_B)}
\left[\kappa \Delta m^2
\begin{pmatrix}
-\cos 2\theta&\sin 2\theta\\
\sin 2\theta& \cos 2\theta\\
\end{pmatrix}
+ \mathcal{A}(r) A[\phi(r)]
\begin{pmatrix}
1&0\\
0&-1\\
\end{pmatrix}
\right]
\begin{pmatrix}
\nu_e\\
\nu_\mu
\end{pmatrix}
,
\end{eqnarray}
where  $\kappa \equiv A^{-1}[\phi(r_B)] A[\phi(r)]$, $\mathcal{A}(r) \equiv 2 \sqrt{2}G_F n_e(r) E_l(r_B) A^{-1}[\phi(r)] \left(1-\frac{\mathcal{R}}{16m_e^2}\right)$ is contained the matter effects, i.e., $\nu_e$-$e^-$ standard interaction, modified in the curved spacetime \cite{Zhang}, and $dl$ is the differential element of the proper distance.
To respect the WEP, we choose various scalar-matter couplings to be the same, i.e., $\beta_1=\beta_2$.
It is also worth noting that the matter effects, from electron-neutrino interactions on neutrino flavor change, increase with the neutrino energy $E_l$.

Hereafter, the critical point is the diagonalization of the effective Hamiltonian and the derivation of the explicit expressions for the effective mixing parameters.
The effective mixing parameters, which are related to the vacuum ones, are given by
\begin{eqnarray}\label{eqn38}
\Delta m^2_M = \sqrt{\left[\Delta m^{2} \sin 2\theta\right]^2 + \left[\Delta m^{2} \cos 2 \theta - \mathcal{A}(r) A[\phi(r_B)]\right]^2},
\end{eqnarray}
and
\begin{eqnarray}\label{eqn39}
\sin 2\theta_{M} = \frac{\Delta m^{2} \sin 2\theta}{\sqrt{\left[\Delta m^{2} \sin 2\theta\right]^2 + \left[\Delta m^{2} \cos 2 \theta - \mathcal{A}(r) A[\phi(r_B)]\right]^2}}.
\end{eqnarray}
Hence, we can treat the  neutrino flavor conversion in the matter as the vacuum neutrino oscillations provided that we use these mass and mixing parameters.

Before introducing our numerical analysis of the solar neutrino data in the next section, it is worth discussing the analytical formula for neutrino survival probability inside matter.
The neutrino state corresponding to flavor $\alpha$ can be suggested as follows:
\begin{eqnarray}\label{eqn40}
\ket{\upnu(\theta,r,t)}_\alpha = \sum_{i,\beta} e^{-i\varphi_i} U_{\beta i}(\theta,r) U_{\alpha i}^*(\theta_M , r_0) \ket{\nu_\beta},
\end{eqnarray}
where $U_{\alpha i}(\theta_M , r_0)$ and $U_{\beta i}(\theta,r)$ are the mixing matrix elements in the matter at the production point and in the vacuum at the detection point.
By averaging out the phase part, the two-flavor electron-neutrino survival probability is then given by
\begin{eqnarray}\label{eqn41}
P_{ee}(E_l) = \frac{1}{2}\left[1 + \cos 2\theta \cos 2\theta_{M} \right].
\end{eqnarray}

\section{Results, discussions and conclusion}\label{sec5}
Inspired by the models of neutrino oscillations in curved spacetime and the screening mechanisms, we have investigated the possible effects of a conformally coupled scalar field on the flavor change process both in the vacuum and in the matter.
We considered a model in which spacetime is filled with a scalar field, e.g., Chameleon or Symmetron, and the neutrino oscillations were considered.
To derive the flavor oscillation phases and probabilities in the vacuum, we solved the Dirac equation by using an approximate method, called the WKB approach (see sec. \ref{sec2}).
We also studied briefly the effects of neutrino forward elastic scattering from electrons in matter, the MSW effect, in section \ref{sec4}.
The presence of the scalar field modifies the neutrino survival probability $P_{ee}(E_l)$ via the LMA-MSW effect as the neutrinos propagate through the slowly varying density of matter.

To study the behavior of the scalar field, we first expand the scalar field around its background value and obtain the equations for the fluctuations inside and outside a nearly spherically symmetric star with the matter density profile depicted in Fig. \ref{fig5}.
The figures are plotted for both the Chameleon and Symmetron fields in Figs.\ref{fig6} and \ref{fig7}.

\begin{figure}[H]
	\centering
	\includegraphics[scale=0.53]{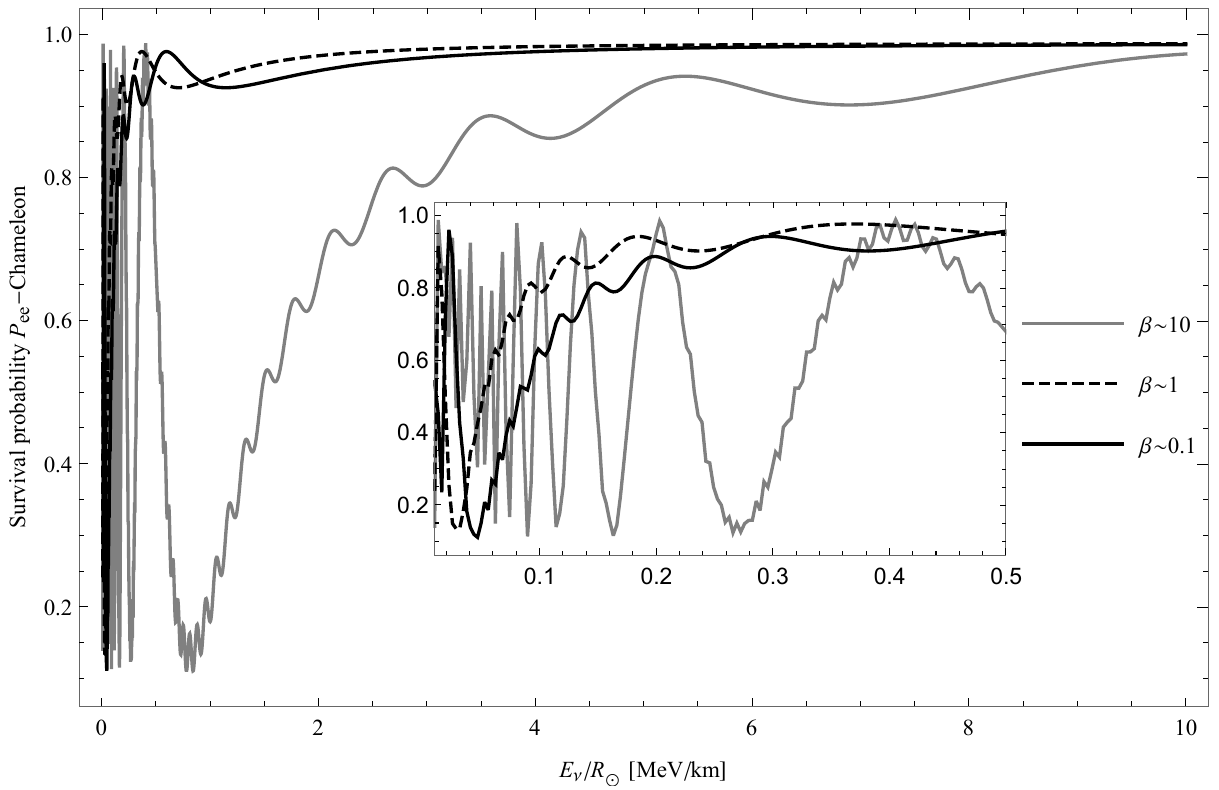}
	\caption{The Solar electron-neutrino survival probability in terms of ratio $E_{\nu}/R_{\odot}$, for $n=1$ and $M=2.08$keV.
		Different curves describe the $P_{ee}$ for various values of the coupling parameter $\beta$.
		Rapid oscillations by increasing $\beta$ result from the Chameleon-dependent oscillation phase.}
	\label{fig1}
\end{figure}

Now let us illuminate the results via some numerical examples.
In Fig.\ref{fig1}, we plot the survival probability $P_{ee}$ as a function of (neutrino energy/solar radius) for the three-flavor case.
We have picked the numerical values $n=1$, and $M \simeq 2.08$keV for Chameleon\cite{Waterhouse:2006wv}, and $\tan^2 \theta_{12}=0.41$, $\sin^2 2\theta_{23} \simeq 0.99$, $\sin^22\theta_{13} \simeq 0.09$,
\[ \Delta m_{21}^2=7.4 \times 10^{-5} eV^2 \] and \[|\Delta m_{23}^2|=2.5 \times 10^{-3} eV^2\]
for Solar neutrinos mixing parameters in 3$\nu$ case \cite{Esteban:2020cvm}.
We have also picked three values for coupling strength $\beta$.
In the Chameleon model proposed by Khoury and Weltman \cite{Khoury:2003aq}, the scalar field might be coupled to matter with $\beta \sim \mathcal{O}(1)$ in agreement with expectations of the string theory \cite{Mota:2006fz}, or smaller, i.e., $\beta \ll 1$.
Setting $\beta \sim 0$ signifies the no-Chameleon case. We have plotted figure \ref{fig1} for $\beta \sim \{0.1,1,10\}$ \cite{Mota:2006fz,Touboul,Mester:2001,Nobili:2000}.

\begin{figure}[H]
	\centering
	\includegraphics[scale=0.6]{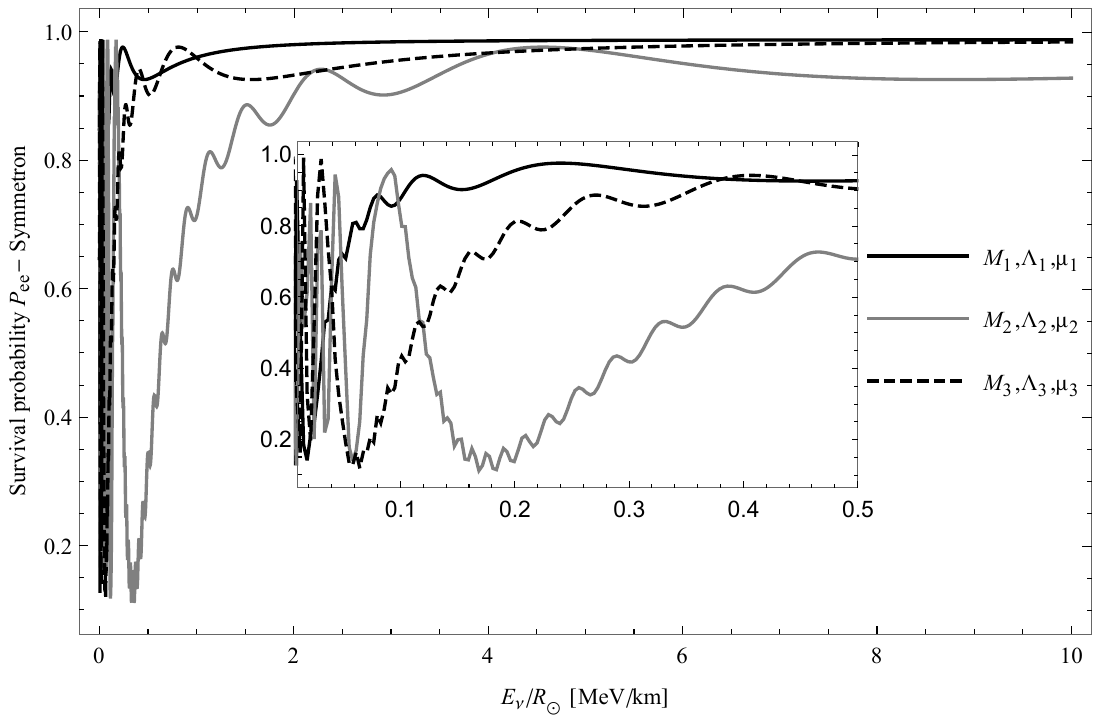}
	\caption{This figure describes the behavior of the $P_{ee}$ in the presence of the Symmetron scalar field, for three various values $\{M_1,M_2,M_3\} \in \{10^{-4}, 10^{-5}, 10^{-6}\} M_p$, $\{\Lambda_1,\Lambda_2,\Lambda_3\} \in \{10^{-64}, 10^{-67}, 10^{-70}\}$, and $\{\mu_1,\mu_2,\mu_3\} \in \{10^{-10}, 10^{-12}, 10^{-14}\}$eV.}
	\label{fig2}
\end{figure}

Also, the survival probability $P_{ee}$ and its oscillating behavior affected by the Symmetron field is shown in Fig.\ref{fig2} for three sets of values of $\{M_i,\Lambda_i,\mu_i\}$.
We have used the constraints imposed by the local tests of gravity on the parameters: $M \leq 10^{-4} M_p$ \cite{Hinterbichler:2011ca,Hinterbichler:2010es}, $\lambda \gtrsim 10^{-96}$ \cite{Hinterbichler:2011ca}.

\begin{figure}[H]
	\centering
	\includegraphics[scale=0.6]{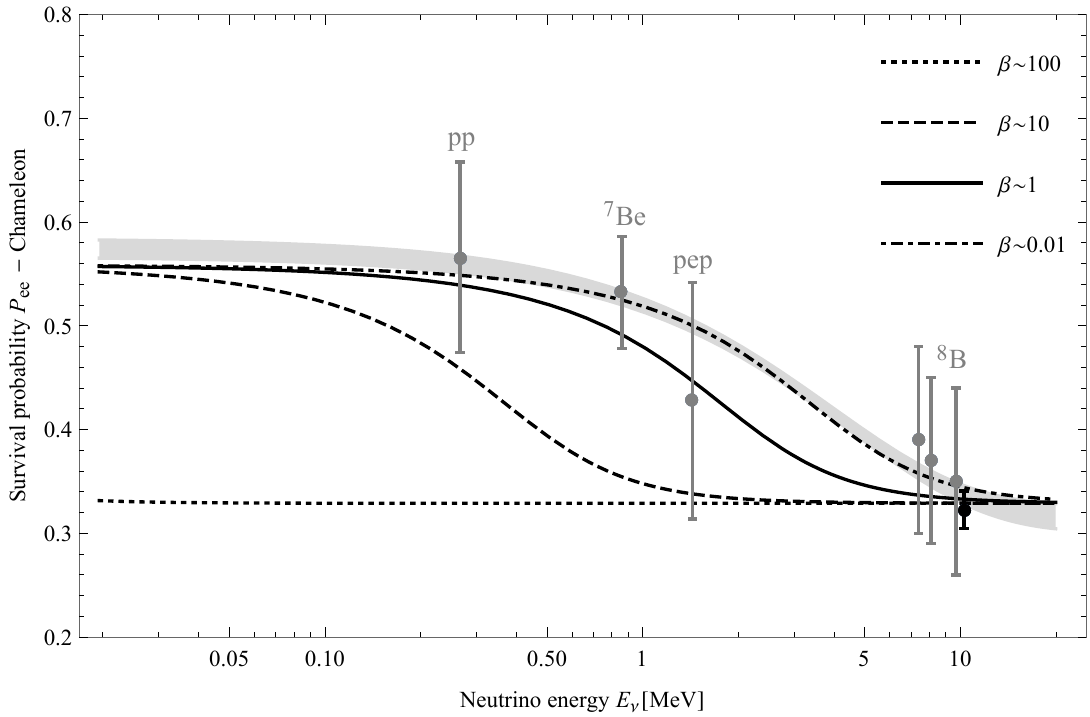}
	\caption{Electron-neutrino survival probability as a function of its energy in [MeV].
	This figure describes $P_{ee}$ including the MSW-LMA effect in the presence of the Chameleon scalar field for different $\beta$'s.
	The gray band illustrates the theoretical prediction of $\nu_e$ survival probability by standard interactions in a flat spacetime.}
	\label{fig3}
\end{figure}

\begin{figure}[H]
	\centering
	\includegraphics[scale=0.58]{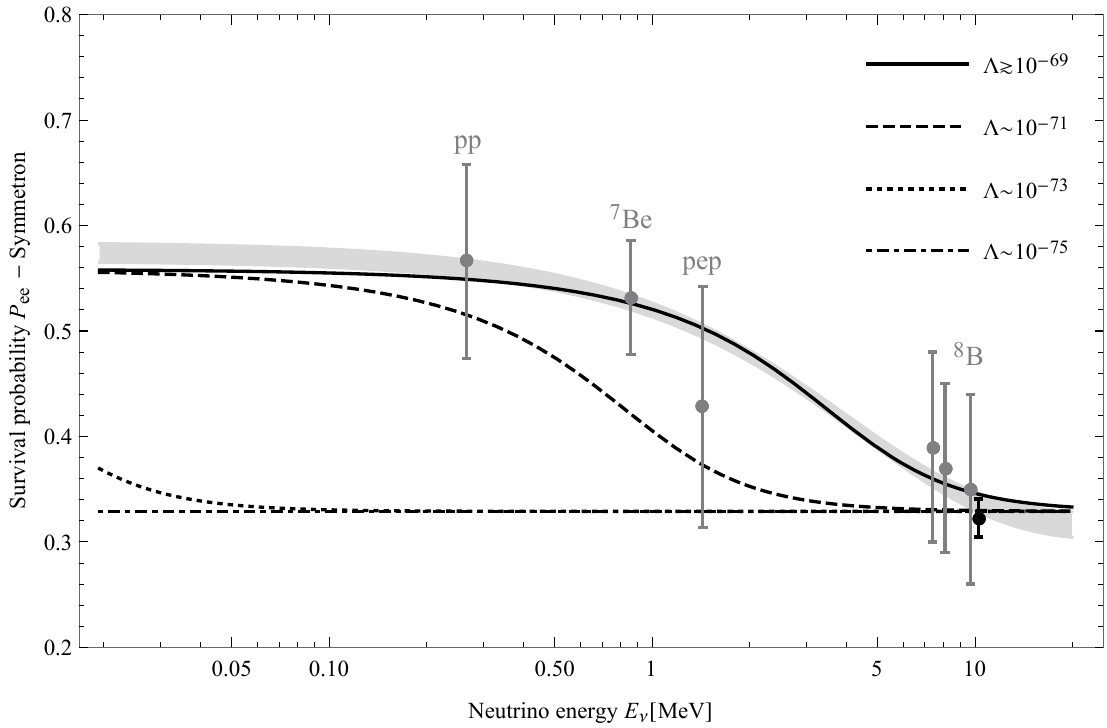}
	\caption{This plot shows the effect of the MSW-LMA scenario on $P_{ee}$ in the presence of the Symmetron scalar field for different values of dimensionless parameter $\Lambda$.
		As before, the gray band in this panel comes from the MSW-LMA theoretical prediction of standard interactions within $\pm 1 \sigma$.}
	\label{fig4}
\end{figure}

Figures \ref{fig3} and \ref{fig4} illustrate the effect of MSW-LMA on the $\nu_e$ survival probability for several values of the scalar field parameters.
We have also used the $P_{ee}$-experimental values from the Borexino data \cite{Agostini:2018uly} of pp, $^7$Be, pep, and $^8$B fluxes (gray points).
Black point also represents the SNO + SK $^8$B data \cite{Zyla:2020zbs}.
The gray band in both plots is the best theoretical prediction of $P_{ee}$ (within $\pm 1 \sigma$) according to the MSW-LMA solution \cite{Agostini:2018uly}.

According to Eq.(\ref{eqn39}), the matter effect potential dominates for the Chameleon-matter coupling $\beta \gtrsim 10^2$, i.e., $\mathcal{A}(r) A[\phi(r_B)] > \Delta m^2 \cos2\theta$, and the survival probability tends to $P_{ee} \sim \sin^2 \theta \sim 0.33$.
For weaker couplings, the effects of the scalar field in the metric and also in matter potential are negligible, and we are left with the MSW-LMA in the flat spacetime.

The electron-neutrino survival probability shown in Fig.\ref{fig4} is in good agreement with observational data in the high-energy range in the presence of the Symmetron scalar field.
This figure is plotted for $M = 10^{-4}M_p$, $\mu \sim 10^{-12}$eV and different values of dimensionless parameter $\Lambda$.
By decreasing $\Lambda$ in orders of magnitude, according to the Eq.(\ref{eqn39}), this figure shows $\mathcal{A}(r) A[\phi(r_B)] > \Delta m^2 \cos2\theta$, and consequently, inconsistency in lower energies.

\appendix
\numberwithin{equation}{section}
\makeatletter
\newcommand{\section@cntformat}{Appendix \thesection:\ }
\makeatother

\section{Screening mechanism}\label{app1}
In the appendix, we will review Chameleon and Symmetron screening models for two cases inside and outside an object, e.g., the Sun, whose density distribution function is approximated by an exponential function of the form $\rho_{\odot}(R)=\rho_c \exp[-\lambda R_{\odot} R]$, see Fig.\ref{fig5}:

\begin{figure}[H]
	\centering
	\includegraphics[scale=1]{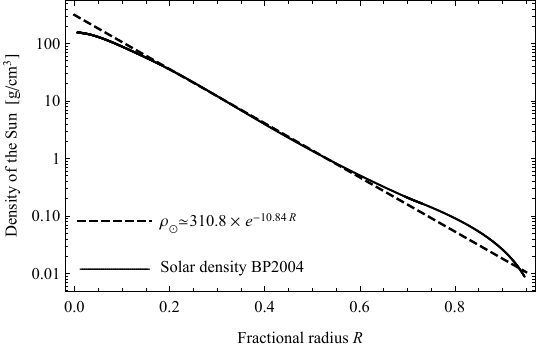}
	\caption{The Solar matter density profile is drawn in terms of the fractional radius $R \equiv \frac{r}{R_{\odot}}$ for the BP04 model \cite{Bahcall:2004fg} (solid line) and the exponential approximation (dashed line).}
	\label{fig5}
\end{figure}
Variation of the action (\ref{eqn1}) with respect to the scalar field $\phi$ gives the equation of motion
\begin{eqnarray}\label{eqnA1}
\square \phi = V_{,\phi} + A_{,\phi} \rho,
\end{eqnarray}
where $\rho$ is the matter density.
For future considerations, it is more convenient to write the right-hand side as the first derivative of an effective potential for the scalar field.

To solve the equation (\ref{eqnA1}) in a spherically symmetric static background, we assume
\begin{eqnarray}\label{eqnA2}
\begin{split}
& \frac{d\phi}{dr} = 0  ~~~~~~~~~~~\text{at}~~~ r\longrightarrow 0,
\\& \phi \longrightarrow \phi_0 ~~~~~~~~~\text{at}~~~ r\longrightarrow \infty.
\end{split}
\end{eqnarray}
The first condition is for the non-singularity of the scalar field at the center of the spherically symmetric body.
The second one implies that the field converges at infinity.

\numberwithin{equation}{subsection}
\subsection{Chameleon mechanism}\label{subapp1}
This model is specified by a potential function of continuously decreasing form
\begin{eqnarray}\label{eqnA1.1}
V(\phi)= M^{4+n}\phi^{-n},
\end{eqnarray}
and an exponential coupling function
\begin{eqnarray}\label{eqnA1.2}
A(\phi) = \exp\left[{\frac{\beta\phi}{M_p}}\right],
\end{eqnarray}
where $n$ is a positive number, $M$ is a parameter of mass scale, and $\beta$ is the coupling strength.

The solution to the Eq.(\ref{eqnA1}) can be obtained by expanding the field about its background as $\phi(r) = \phi_0 + \delta \phi$ up to linear order, where $\phi_0$ is the constant background value and $\delta \phi$ is the perturbation induced by a spherically symmetric body like the Sun, whose radius is $R_{\odot}$.
Therefore the field equation turns into
\begin{eqnarray}\label{eqnA1.3}
\frac{d^2 \delta\phi}{dr^2} + \frac{2}{r} \frac{d\delta\phi}{dr} \simeq m_{\text{min}}^2 (\phi_0) \delta\phi + \frac{\beta(\phi_0)}{M_p} \rho(r).
\end{eqnarray}
Because the cosmological and local gravity experiments impose the condition $\phi/M_p \ll 1$, we assume that $e^{\frac{\beta \phi}{M_p}} \approx 1 + \beta \phi/M_p$.
Also, we have assumed that the Schwarzschild radius of the spherical object is small enough such that $G m(r)/r , Gm'(r) \ll 1$.
The quantity
\begin{eqnarray}\label{eqnA1.4}
\begin{split}
&m^2_{\text{min}} = \frac{n(n+1) M^{4+n}}{\phi^{n+2}_{\text{min}}} + \frac{\rho \beta^2}{M_p^2}
\end{split}
\end{eqnarray}
gives the effective mass of the Chameleon field, where the minimum of the potential is determined by using the equation $V_{\text{eff},\phi}(\phi_{\text{min}}) = 0$, which leads to
\begin{eqnarray}\label{eqnA1.5}
\phi_{\text{min}} = \bigg[\frac{n M^{4+n} M_p}{\beta \rho}\bigg]^{\frac{1}{n+1}}.
\end{eqnarray}
The main feature of the Chameleon scalar field is that its effective mass depends explicitly on the matter density.

We can find an analytical solution to the scalar field equation (\ref{eqnA1.3}) inside and outside the Sun with a constant background matter density $\rho_0$. We have
\begin{eqnarray}\label{eqnA1.6}
\begin{split}
&\delta\phi_{\text{in}}(R) = \frac{e^{-(\lambda + 2m_{\text{in}})R_{\odot} R}}{2 m_{\text{in}} M_p R_{\odot} \left(m_{\text{in}}^2 - \lambda^2\right)^2 R} \bigg[C_1 M_p \left(m_{\text{in}}^2 - \lambda^2 \right)^2 \left(e^{2 m_{\text{in}} R_{\odot} R} - 1 \right) e^{(\lambda + m_{\text{in}}) R_{\odot} R} \\& ~~~~~~~~~~~+ 2 \beta  m_{\text{in}}  \rho_c  \left(e^{2 m_{\text{in}} R_{\odot} R} \left(2 \lambda + \left(\lambda^2 - m_{\text{in}}^2\right) R_{\odot} R\right) - 2 \lambda e^{(\lambda + m_{\text{in}}) R_{\odot}R}\right)\bigg], &&(R<1)
\end{split}
\end{eqnarray}
and
\begin{eqnarray}\label{eqnA1.7}
\begin{split}
\delta\phi_{\text{out}}(R) = C_2 \frac{e^{- m_{\text{out}} R_{\odot} R}}{R}, && &&(R>1)
\end{split}
\end{eqnarray}
where $C_1$ and $C_2$ are both constants of integration obtained from continuity conditions of the scalar field and its first derivative at the boundary $R=1$.
By assuming $m_{\text{in}} R_{\odot} \gg 1$ and $m_{\text{out}} R_{\odot} \ll 1$, which result in the screening inside the object and a long-range force outside it, we fix these two constants as follows:
\begin{eqnarray}\label{eqnA1.8}
\begin{split}
&C_1 = \frac{2 \beta  \rho_c e^{- 2 m_{\text{in}} R_{\odot}}}{M_p (m_{\text{in}}^2 -\lambda^2)^2} \bigg[e^{(m_{\text{in}} -\lambda)R_{\odot}} \left(m_{\text{in}}^2 (1 - \lambda  R_{\odot}) + \lambda \left(\lambda - m_{\text{out}} (\lambda R_{\odot} + 2) + \lambda^2 R_{\odot}\right)\right) - 2 \lambda m_{\text{in}}\bigg],
\end{split}
\end{eqnarray}
and
\begin{eqnarray}\label{eqnA1.9}
\begin{split}
&C_2 = \frac{\beta \rho_c e^{-(2m_{\text{in}} + \lambda)R_{\odot}}} {M_p m_{\text{in}} R_{\odot} (m_{\text{in}}^2 - \lambda^2)^2}   \bigg[ -m_{\text{in}}^3 R_{\odot} e^{2 m_{\text{in}} R_{\odot}} - m_{\text{in}}^2 (\lambda R_{\odot} - 1) e^{2 m_{\text{in}} R_{\odot}} \\& ~~~+ \lambda^2 (\lambda R_{\odot} + 1) e^{2 m_{\text{in}} R_{\odot}} + \lambda m_{\text{in}} \left((\lambda R_{\odot} + 2) e^{2 m_{\text{in}} R_{\odot}} - 4 e^{(\lambda + m_{\text{in}}) R_{\odot}}\right)\bigg].
\end{split}
\end{eqnarray}
Adding these solutions by the background value $\phi_0$, we obtain the whole profile of the scalar field.
As can be seen from these relations, the scalar field induced by the Sun is too tiny in large distances such that we can ignore its effects on another object in the solar system scale.
As a special case, by setting $\lambda \rightarrow 0$, all above relations can be re-written for constant or slowly varying matter density $\rho_c$.

\begin{figure}[H]
	\centering
	\includegraphics[scale=0.57]{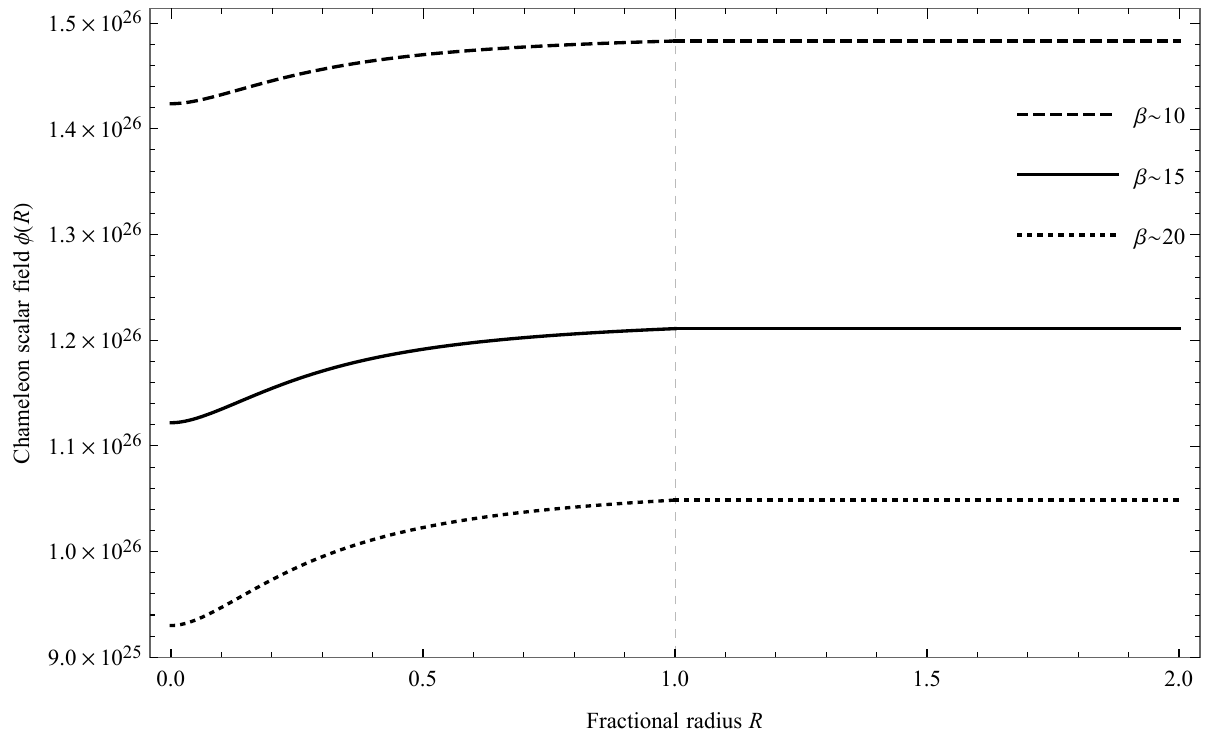}
	\caption{These curves show how the coupling parameter might affect the Chameleon inside and outside the Sun.
		The Chameleon, in each case, approaches an asymptotic value outside the body, which increases with decreasing $\beta$.
		The field values are all in eV.}
	\label{fig6}
\end{figure}

Figure \ref{fig6} shows how the Chameleon scalar field is affected by three different coupling parameters $\beta \in \{10,15,20\}$, implying that the Chameleon tends to smaller asymptotic values when $\beta$ increases.
We note that the field's allowed range becomes smaller when $\beta$ declines.

\numberwithin{equation}{subsection}
\subsection{Symmetron mechanism}\label{subapp2}
The Symmetron screening process works by symmetry restoration in the dense regions in which $\phi = 0$ where the coupling between Symmetron and matter leads to zero.
However, in low-density regions, the $\mathbb{Z}_2$-symmetry is spontaneously broken, and the vacuum expectation value (VEV) of the scalar field does not vanish.
The coupling function and potential are chosen such that they obey the $\mathbb{Z}_2$-symmetry ($\phi \rightarrow -\phi$ symmetry):
\begin{eqnarray}\label{eqnA2.1}
\begin{split}
&A(\phi) = 1 + \frac{1}{2M^2} \phi^2+ \mathcal{O}\bigg(\frac{\phi^4}{M^4}\bigg),
\\&V(\phi) = V_0 - \frac{1}{2} \mu^2 \phi^2 + \frac{1}{4} \Lambda \phi^4,
\end{split}
\end{eqnarray}
where $M$ and $\mu$ are two parameters of mass scale, and $\Lambda$ is a dimensionless parameter.
The equation of motion of the scalar field in the static spherically symmetric background is given by
\begin{eqnarray}\label{eqnA2.2}
\square \phi = \frac{1}{M^2} \left(\rho(r) - \uprho_C \right) \phi + \Lambda \phi^3,
\end{eqnarray}
where $\uprho_C \equiv \mu^2 M^2$ is the critical density.
The breaking or restoration of $\mathbb{Z}_2$-symmetry depends on whether the matter density is smaller or larger than the critical density.
In the dilute regions with $\rho_0 \ll \uprho_C$, the symmetry is broken, and the scalar field acquires a VEV
\begin{eqnarray}\label{eqnA2.3}
\phi_{\text{min,out}} = \pm \frac{\mu}{\sqrt{\Lambda}} \sqrt{1-\frac{\rho_0}{\uprho_C}} \approx \pm \frac{\mu}{\sqrt{\Lambda}}. && (\rho_0 \ll \uprho_C)
\end{eqnarray}
For the Sun, the scalar field's effective mass is then
\begin{eqnarray}\label{eqnA2.4}
\begin{split}
& m_{\text{out}} = \sqrt{2}\mu \sqrt{1 - \frac{\rho_{0}}{\uprho_C}} \approx \sqrt{2} \mu, && (R>1)
\\& m_{\text{in}}(r) = \mu \sqrt{\frac{\rho_{\odot}(r)}{\uprho_C} - 1} \approx \mu \sqrt{\frac{\rho_{\odot}(r)}{\uprho_C}}, &&(R<1)
\end{split}
\end{eqnarray}
where, as before, $\rho_0$ is the background matter density, and $\rho_{\odot}(r)$ is the solar density distribution function, depicted in Fig. \ref{fig5}.

By expanding the scalar field around its background value $\phi_0 \equiv \phi_{\text{min,out}}$ and assuming that the Schwarzschild radius of the compact object is small enough such that $G m(r)/r \ll 1$ and $G m'(r) \ll 1$, we find the equation of motion as follows:
\begin{eqnarray}\label{eqnA2.5}
\frac{d^2 \delta\phi}{dr^2} + \frac{2}{r} \frac{d\delta\phi}{dr} \simeq m_{\text{min}}^2(\phi_0) \delta\phi + \frac{\phi_0}{M^2} \rho(r).
\end{eqnarray}
Assuming $\rho_0 \ll \bar{\rho}_{\odot}$ for the background density outside the Sun, the solution to this equation is given by
\begin{eqnarray}\label{eqnA2.6}
\delta\phi_{\text{out}}(R) = C_1 \frac{e^{-m_{\text{out}} R_{\odot} (R-1)}}{R}. && (R>1)
\end{eqnarray}
The constant $C_1$ is determined by continuity conditions at $R=1$:
\begin{eqnarray}\label{eqnA2.7}
\begin{split}
& C_1 = -\phi_0 \biggl[K_0\left\{2 \sqrt{\frac{\rho_c}{M^2 \lambda ^2}}\right\} I_0\left\{2 \sqrt{\frac{e^{-R_{\odot} \lambda } \rho_c}{M^2 \lambda ^2}}\right\} - I_0\left\{2 \sqrt{\frac{\rho_c}{M^2 \lambda^2}}\right\} K_0\left\{2 \sqrt{\frac{e^{-R_{\odot} \lambda} \rho_c}{M^2 \lambda^2}}\right\}\\& + R_{\odot} \sqrt{\frac{\rho_c e^{-\lambda R_{\odot}}}{M^2}} \left(K_0\left\{2 \sqrt{\frac{\rho_c}{M^2 \lambda^2}}\right\} I_1\left\{2 \sqrt{\frac{e^{- \lambda R_{\odot}} \rho_c}{M^2 \lambda ^2}}\right\} - I_0\left\{2 \sqrt{\frac{\rho_c}{M^2 \lambda^2}}\right\} K_1\left\{2 \sqrt{\frac{e^{-\lambda R_{\odot}} \rho_c}{M^2 \lambda ^2}}\right\}\right) \biggl]     /     \\& \left[R_{\odot} \sqrt{\frac{\rho_c e^{-\lambda R_{\odot}}}{M^2}} \left(K_0\left\{2 \sqrt{\frac{\rho_c}{M^2 \lambda^2}}\right\} I_1\left\{2 \sqrt{\frac{e^{-\lambda R_{\odot}} \rho_c}{M^2 \lambda ^2}}\right\} +  I_0\left\{2 \sqrt{\frac{\rho_c}{M^2 \lambda^2}}\right\} K_1\left\{2 \sqrt{\frac{e^{-\lambda R_{\odot}} \rho_c}{M^2 \lambda^2}}\right\}\right)\right],
\end{split}
\end{eqnarray}
where $I_n(z)$ and $K_n(z)$ are modified Bessel functions of the first and second kind, respectively.
Note that we have assumed that $m_{\text{out}} R_{\odot} \ll 1$, which implies on a long-range force outside the spherical object.

\begin{figure}[H]
		\centering
		\includegraphics[scale=0.52]{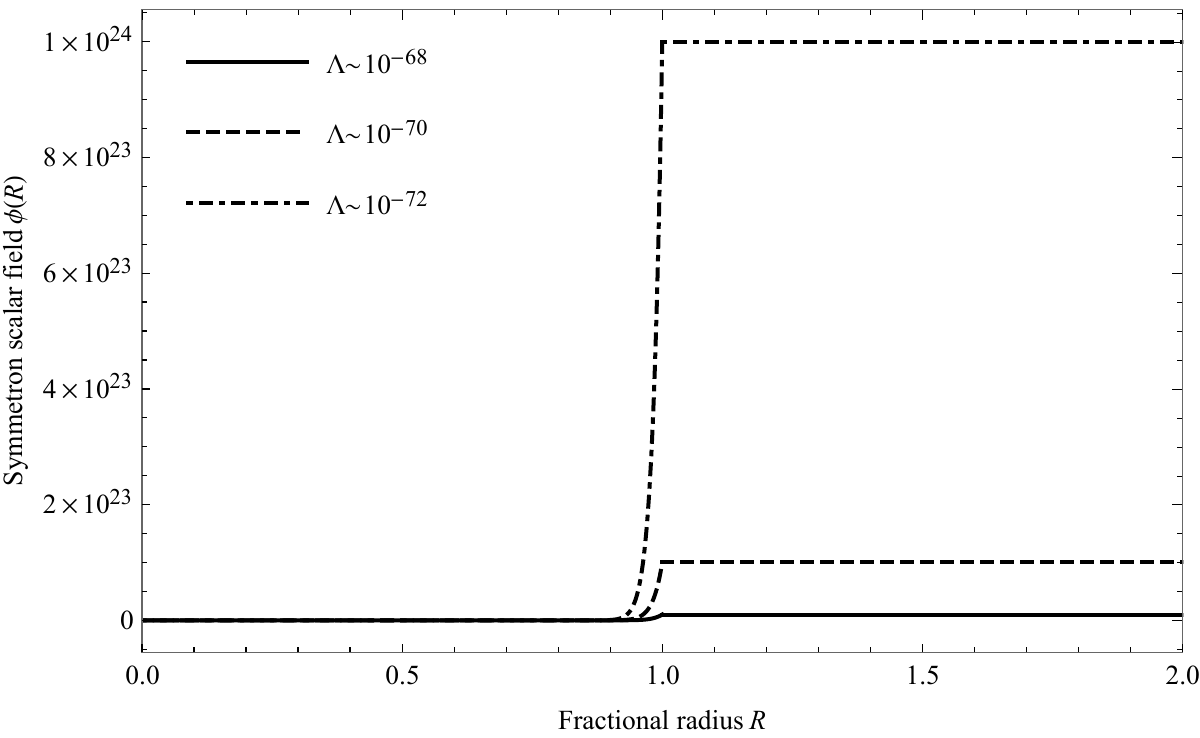}
	\caption{This plot shows the effects of the parameter $\Lambda$ on the Symmetron field for $M \sim 10^{-4}M_p$, $\mu\sim 10^{-12}$eV.}
		\label{fig7}
\end{figure}

In a dense region, where $\rho > \uprho_C$, we obtain the equation of motion inside the Sun as follows:
\begin{eqnarray}\label{eqnA2.8}
\frac{d^2 \delta\phi}{dr^2} + \frac{2}{r} \frac{d\delta\phi}{dr} = m_{\text{in}}^2(r) \left(\delta\phi + \phi_0\right). && (r<R_{\odot})
\end{eqnarray}
The solution to this equation is
\begin{eqnarray}\label{eqnA2.9}
\delta\phi_{\text{in}}(R) = -\sqrt{2} C_2  \frac{K_0\left\{2 \sqrt{\frac{e^{-\lambda R_{\odot} R} \rho_c}{M^2 \lambda ^2}}\right\}}{\lambda R_{\odot}R} + \frac{\sqrt{2} C_2 K_0 \left\{2 \sqrt{\frac{\rho_c}{M^2 \lambda^2}}\right\}}{I_0\left\{2 \sqrt{\frac{\rho_c}{M^2 \lambda^2}}\right\}} \frac{I_0\left\{2 \sqrt{\frac{e^{-\lambda R_{\odot} R} \rho_c}{M^2 \lambda^2}}\right\}}{\lambda R_{\odot} R} , & (R<1)
\end{eqnarray}
where
\begin{eqnarray}\label{eqnA2.10}
\begin{split}
& C_2 = - \frac{\lambda \phi_0 M}{\sqrt{2 \rho_c e^{-\lambda R_{\odot}}}} \frac{I_0\left\{2 \sqrt{\frac{\rho_c}{M^2 \lambda ^2}}\right\}}{I_0\left\{2 \sqrt{\frac{\rho_c}{M^2 \lambda^2}}\right\} K_1\left\{2 \sqrt{\frac{e^{-\lambda R_{\odot}} \rho_c}{M^2 \lambda^2}}\right\} +  K_0\left\{2 \sqrt{\frac{\rho_c}{M^2 \lambda^2}}\right\} I_1\left\{2 \sqrt{\frac{e^{-\lambda R_{\odot}} \rho_c}{M^2 \lambda^2}}\right\}}.
\end{split}
\end{eqnarray}
Figure \ref{fig7} is plotted to show how the dimensionless parameter $\Lambda$ affects the Symmetron scalar field.
All curves are depicted for the numerical values $\mu \sim 10^{-12}$eV, $M \sim 10^{-4} M_p$, and $\Lambda \in \{10^{-68}, 10^{-70}, 10^{-72}\}$.
As in the Chameleon case, the Symmetron tends to an asymptotic value $\phi_{0}$ at large distances from the Sun.

\end{document}